\documentclass[preprint]{aastex}
\usepackage{graphicx}
\usepackage{amssymb}
\usepackage{amsmath}
\newcommand{\ls}{LS\,5039}
\newcommand{\lsi}{LS\,I\,+61\,303}

\begin{document}

\title{Extended X-ray emission in the vicinity of the microquasar LS 5039: pulsar wind nebula?}
\author{Martin Durant$^1$, Oleg Kargaltsev$^1$, George G. Pavlov$^{2,3}$, Chulhoon Chang$^2$,\\ Gordon P. Garmire$^2$}
\affil{$^1$ University of Florida, 211 Bryant Space Science Center, Gainesville, FL, USA\\
$^2$ Pennsylvania State University, 525 Davey Lab, University Park, PA, USA\\
$^3$ St.-Petersburg State Polytechnical University, Polytekhnicheskaya ul.\ 29, St.-Petersburg,
195251, Russia}
\email{martin.durant@astro.ufl.edu}

\begin{abstract}
\ls\ is a high-mass binary with a period of 4 days, containing a compact object and an O star, one of the few high-mass binaries detected in $\gamma$-rays. Our Chandra ACIS observation of LS 5039 provided a high-significance ($\approx10\sigma$) detection of  extended emission clearly visible for up to 1\arcmin\ from the point source.  The spectrum of this emission can be described by an absorbed power-law model with photon index $\Gamma=1.9\pm0.3$, somewhat softer than the point source spectrum $\Gamma=1.44\pm0.07$, with the same absorption, $N_H=(6.4\pm0.6)\times10^{21}$\,cm$^{-2}$.  The observed 0.5--8\,keV flux of the extended emission is $\simeq8.8\times10^{-14}$\,erg\,s$^{-1}$cm$^{-2}$, or  5\% of the point source flux; the latter is a factor of $\sim2$ lower than the lowest flux detected so far. Fainter extended emission with comparable flux and a softer ($\Gamma\approx3$) spectrum is detected at even greater radii (up to 2\arcmin).
Two possible interpretations of the extended emission are a dust scattering halo and a synchrotron nebula powered by energetic particles escaping the binary. We discuss both of these scenarios and favor the nebula interpretation, although some dust contribution is possible.  We have also found transient sources located within a  narrow stripe south of LS 5039. We discuss the likelihood of these sources to be related to \ls.

\end{abstract}
\maketitle

\section{Introduction}

In a high-mass X-ray binary (HMXB), 
a compact object,
which can be  a black hole (BH) or a neutron star (NS),
accretes some portion of the wind from  a high-mass ($M>5M_\odot$)
 non-degenerate  companion star.
Some HMXBs are called `micro-quasars' ($\mu$QSOs) when resolved, collimated radio features are observed, which are interpreted as relativistic jets.

There has been renewed interest in $\mu$QSOs since the detection of parsec-scale X-ray jets \citep{2007RMxAC..27..122C} and variable   $\gamma$-ray emission \citep{2009Sci...326.1512F}. The nature of the high-energy emission for two micro-quasars, \ls\ and \lsi, has been much debated, e.g., the emission could truly be
generated in a $\mu$QSO jet \citep{2008MNRAS.383..467K}, or possibly 
it could be generated by the interaction of a young pulsar wind with the wind of the companion \citep{2004A&A...413.1019L,2006smqw.confE..52D,2006A&A...456..801D}.

\ls\ was first identified as an HMXB by \citet{1997A&A...323..853M} based on the positional coincidence of an X-ray source with an O7V star ($V=11.5$), projected near the Galactic bulge ($l=16\fdg8816$, $b=-1\fdg2892$).
Using optical and infra-red spectroscopy,  \citet{2001A&A...376..476C} found the companion 
 star to be type O6.5V(f),
variable in the IR but not in the optical. \cite{2001ApJ...558L..43M} measured line radial velocities and found 
  binary period $P_{\rm orb}=4.1$\,d and substantial eccentricity $e=0.4$. Using detailed time-resolved spectroscopy and atmospheric modeling of the companion, \citet{2005MNRAS.364..899C} 
  refined the binary parameters ($P_{\rm orb}=3.906$ d, $e=0.35$) and concluded that the system must harbor a high-mass compact object, probably a BH, for an inferred distance of 
$2.5\pm 0.1$\,kpc. The conclusion was, however, later  disputed by \citet{2006A&A...456..801D}  who argued that the NS option still cannot be dismissed. Using high-precision, long time base-line photometry from the {\sl MOST} satellite and ground-based spectroscopy, \citet{2011MNRAS.411.1293S} further refined the orbital characteristics, preferring a slightly lower eccentricity $e=0.24\pm0.08$ and component masses 26$M_\odot$ for the O-star and 1.7$M_\odot$ for the compact object. Thus, the nature of the compact object in LS 5039 remains unknown.

An unresolved, non-thermal, variable radio source was fist detected by \citet{1998A&A...338L..71M}. Using the high angular resolution of VLBA, the source was resolved, and `jets' observed by \citet{2000Sci...288.2340P}; the source was therefore classified as a $\mu$QSO.
\citet{2002A&A...384..954R}  measured the system proper motion ($\approx$12\,mas\,yr$^{-1}$, using radio and optical astrometry) and, based on its high tangential velocity, called the system `runaway', i.e., with enough velocity to leave the Galactic disk. The proper motion vector can be traced back to the supernova remnant (SNR) G16.8$-$1.1.
Subsequent radio observations by \citet{2008A&A...481...17R} showed that the resolved emission  of LS 5039 varies in intensity, and that the position angle of the elongated features changes.

LS 5039 has been extensively observed in X-rays, mainly to study its variability and spectrum.
The X-ray emission is modulated at the orbital period. The light-curve produced from X-ray data taken in 1999-2007 was found to be remarkably stable  \citep{2009ApJ...697L...1K},  although some aperiodic variability has been reported earlier with {\sl RXTE} \citep{2003A&A...405..285R,1999A&A...347..518R}.
 
Both \ls\ and \lsi\ have been detected at  higher energies. 
 LS 5039 has been detected  with  HESS \citep{2006A&A...460..743A}, 
   {\sl INTEGRAL} \citep{2009A&A...494L..37H} and {\sl Fermi} \citep{2009ApJ...706L..56A},
with a spectrum similar to $\gamma$-ray pulsars. 
The high-energy light curves show modulations at the orbital period, peaking at superior conjunction 
(see \citealt{2009IJMPD..18..347B} for a review).

In this paper we report 
the results of a 38\,ks  imaging observation 
 of \ls\ with {\em Chandra}.
  In Section \ref{ana}, we describe our data and reduction procedures, as well as several archival data sets. In Section \ref{res} we analyze the spectra of both the point source and the extended emission, calculate radial profiles, and examine the extended morphology. 
  We also describe the properties of a number of transient sources that appear  in the field. 
  In Section \ref{dis} we discuss our findings and their implications. We conclude with a brief summary in Section \ref{sum}.
We also include Appendix A, in which we calculate the X-ray contribution due to dust scattering in the direction of \ls.

\section{Observations and data analysis}\label{ana}
Table \ref{log} lists the 
 X-ray observations considered in this work, including five archival observations (from {\sl XMM-Newton} and {\sl Chandra}) and 
 our latest {\sl Chandra} observation. 

\begin{deluxetable}{ccccc}
\tablecaption{Log of observations \label{log}}
\tabletypesize{\footnotesize}
\tablewidth{0in}
\tablehead{
\colhead{Date} & \colhead{ObsID} & \colhead{Telescope/} & \colhead{Off-axis}  & \colhead{Exposure} \\
& & \colhead{instrument} & \colhead{angle (\arcmin)} & \colhead{ time (ks)}
}
\startdata 
2004-07-09 & 4600 & Chandra/ACIS & 12 & 11.0 \\
2004-07-11 & 5341 & Chandra/ACIS & 12 & 18.0 \\
2005-04-13 & 6259 & Chandra/ACIS & 0 & 5.0\\
2005-09-22 & 0202950201 & XMM/EPIC & 0 & 15.8\\
2005-09-24 & 0202950301 & XMM/EPIC & 0 & 10.4\\
2009-10-30& 10696 & Chandra/ACIS & 3 & 37.8
\enddata
\tablecomments{Exposure times are live (i.e., corrected for GTI filtering and dead-time). }
\end{deluxetable}

\begin{figure*}
\begin{center}
\includegraphics[width=\hsize]{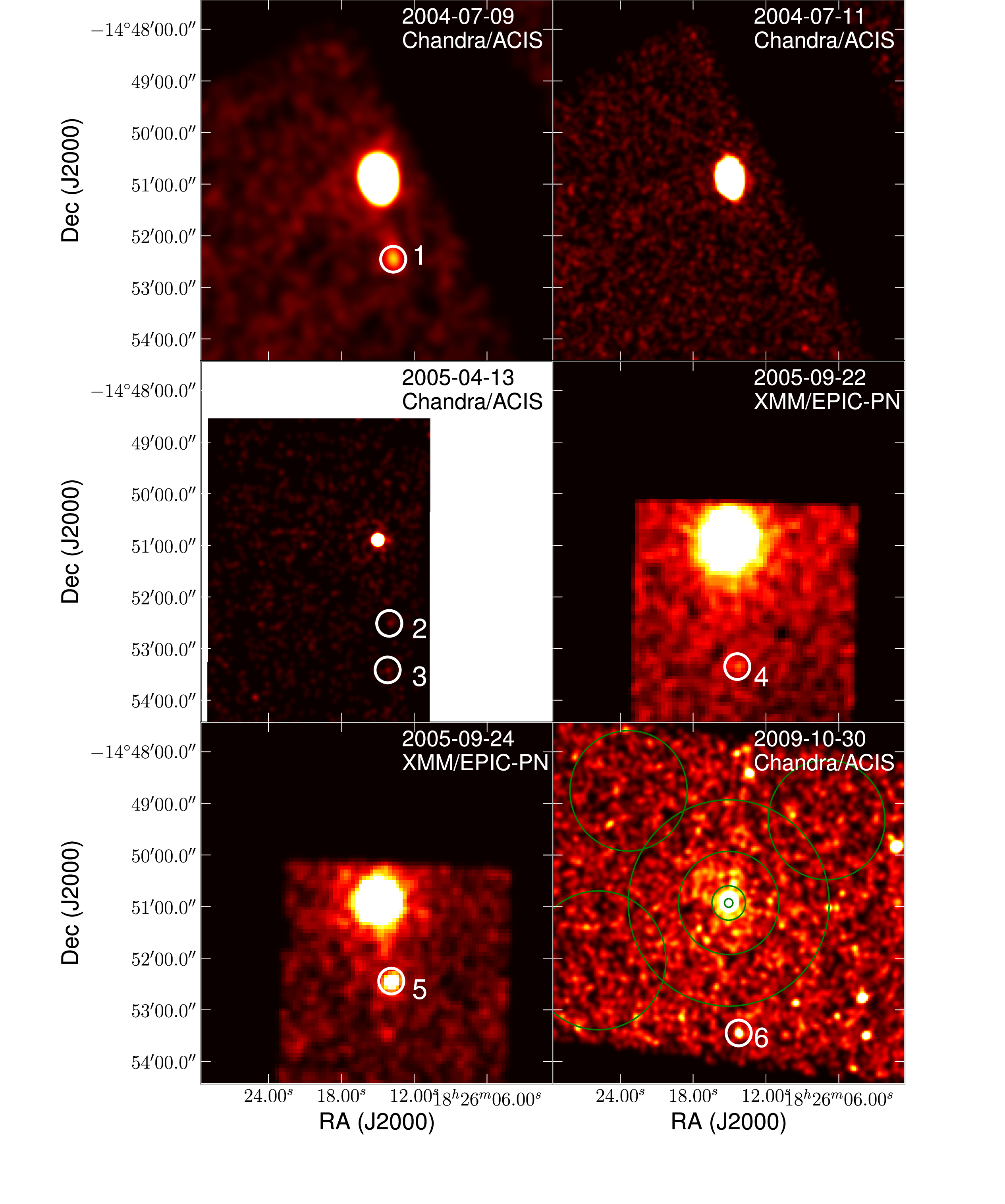}
\caption{Images of the field of \ls.  All images are 7\arcmin$\times$7\arcmin,
in the 0.59--6.8\,keV energy band for {\sl Chandra}, 1--7\,keV for {\sl XMM-Newton}. In the bottom-right panel we mark the source and background extraction regions used in our spectral analysis (green circles; see \S3.2). White numbered circles show the faint field point sources discussed in \S\ref{bob}. }\label{im}
\end{center}
\end{figure*}

\subsection{New Chandra observation}
LS 5039 was observed with the Advanced CCD Imaging Spectrometer (ACIS) 
on board {\sl Chandra} on 2009 October 30 (ObsID 10696). Our observation spanned orbital phases $\phi=0.064$--0.184 (using the
%
%
%
 ephemeris of \citealt{2005MNRAS.364..899C}, where $\phi=0$ corresponds to periastron).
The target was observed for 38.55\,ks 
in timed exposure mode using ``very faint'' telemetry format. 
To minimize  pile-up effects, we used a custom 750 pixel wide sub-array and turned off all the ACIS chips except I3.
The frame time in this configuration is 2.24104\,s (2.2\,s exposure time 
and 0.04104\,s image transfer time).
The Y-offset was $-2\farcm8$, placing the source 3\farcm03 off axis. This  resulted in a broader PSF, further helping to mitigate pile-up. There were no significant  background flares during the observation. The useful scientific exposure time (live-time) was 37.84\,ks. 

The data were reduced and 
analyzed with the {\sl Chandra}\/ Interactive Analysis Observations (CIAO) 
package (ver.\ 4.2), using CALDB 4.2.0.
To produce images of the pulsar and its vicinity at subpixel resolution, 
we removed the pipeline pixel randomization and applied the 
subpixel resolution tool to split-pixel events in the 
image \citep{2001ASPC..251..576M}. 

The image produced from this data is shown as the bottom-right panel in Figure \ref{im}. 

\subsection{Archival observations}
We downloaded archival observations of \ls, three by {\sl Chandra} and two 
by {\sl XMM-Newton}, see Table \ref{log}. In the case of the {\sl Chandra} data, we used the standard Level 2 calibrated science event files. We 
used the energy range 0.59--6.8\,keV, which is the optimal range for viewing the extended emission (see below), and show the resultant images in Figure \ref{im}. 
Although taken with the same instrument, the data sets 
 span a range of exposure times and off-axis distances. 

In the case of the {\sl XMM-Newton} data, we made use of only the EPIC-PN data, from the standard pipeline in the XMM-SAS calibration package. These data were filtered in the energy range 1--7\,keV, and the resultant images are also shown in Figure \ref{im}.

We performed no further processing on any of these archival data, and 
use them only for comparison purposes while looking for transient sources
 (\S \ref{bob}).

In addition to the six observations discussed in this paper, there were two observations by {\sl Chandra} and two by {\sl XMM-Newton}. The {\sl Chandra} data were taken in continuous-clocking mode, and so are not useful for analysis either of extended emission or faint field sources because of the much higher background. For the {\sl XMM} observations, the spacecraft orientation was such that the locations of faint field sources shown in Figure \ref{im} were not imaged, and again the PSF is too broad and the background too high to see extended emission. We consider neither observation further, except to note the point source flux measured by {\sl Chandra}, below.

\section{Results}\label{res}
From the images in Figure \ref{im}, it appears that \ls\ is a bright unresolved point source with extended emission around it; the extended emission
is seen most clearly 
in the latest {\sl Chandra} observation (bottom right panel).
We split our results into three sections: the spectrum of the bright point source, which is essentially unaffected by faint extended emission, the properties of the extended emission, which includes some contribution by the bright source's PSF, and a
 description of the transient sources in the field.

\subsection{Point source}
From the {\sl Chandra} ACIS 
data of 2009 October 30, we extracted 
the spectrum of the 
point-like source corresponding to \ls, centered at  
coordinates\footnote{We use the coordinate system provided by the data processing pipeline.} ${\rm R.A.}=276\fdg56278$, ${\rm decl.}=-14\fdg84871$
(J2000),
 using the CIAO task {\tt psextract}.
 We picked  background regions far from the source, and as free of point sources as possible, yet large enough (total area 51,100 arcsec$^2$)
 that the background spectrum  be well-determined.
The extraction regions are shown on Figure \ref{im}.

\begin{figure*}
\begin{center}
\includegraphics[width=\hsize]{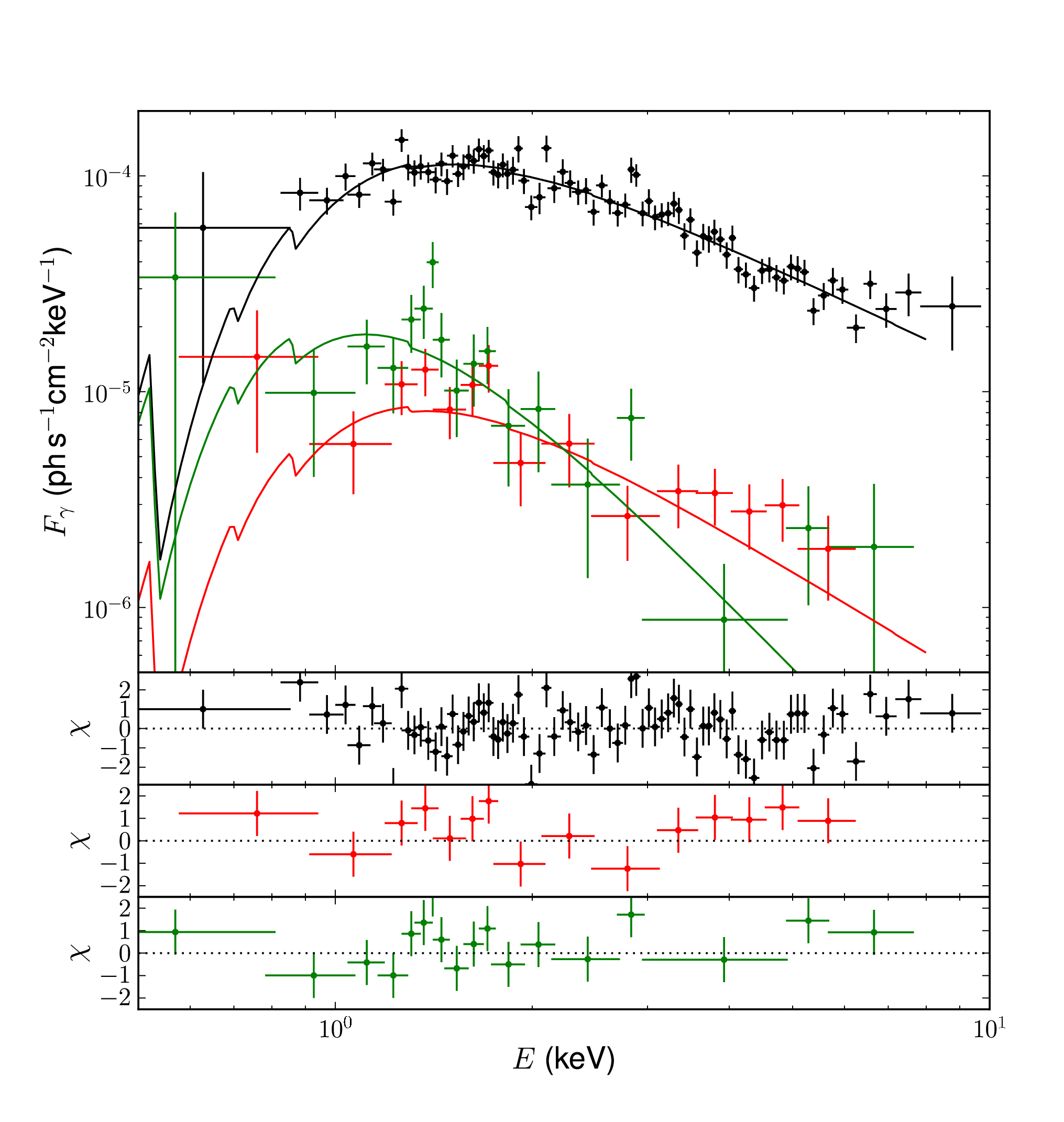}
\caption{Fluxed spectrum of \ls\ (black) and the extended emission extracted from 20\arcsec--60\arcsec\ 
(red) and 60\arcsec--120\arcsec\ (green) annuli around \ls. Best-fit absorbed power-law models are over-plotted as solid lines,  and residuals, $\chi=({\rm data-model)/uncertainty}$ are shown in the lower three panels. The best-fit model
and fitting parameters were estimated using standard forward-fitting
(i.e., by fitting the count-rate spectrum).}\label{fluxed}
\end{center}
\end{figure*}

We fitted an absorbed power-law (PL) model to the spectrum,
extracted from the $r=5\arcsec$ aperture, 
using the {\it Sherpa} package.
We grouped the counts in 25-count bins and fitted in the energy range 
0.5--8\,keV (4602 counts in total, of which 3.8 counts belong to the 
background; 
source count rate 
$0.122\pm 0.002$ counts s$^{-1}$).
 The observed 
energy flux is $F_{\rm 0.5-8\,keV}=(1.92\pm0.03)\times10^{-12}$\,erg\,cm$^{-2}$\,s$^{-1}$. The fitting parameters are given in Table 2, and the fluxed
count-rate spectrum\footnote{The fluxed count-rate spectrum, which we will
call the `fluxed spectrum' for brevity, is calculated by dividing the count 
rate in a given energy bin by an average effective area corresponding to
this bin.} is shown in Figure \ref{fluxed}.

The inferred extinction-corrected flux for the 1--10\,keV range is 2.6$\times10^{-12}$\,erg\,cm$^{-2}$\,s$^{-1}$. Although the phase interval of our observation corresponds to the minimum of the previously observed orbital variability, this is a factor of two lower than any flux that has been noted in the past (e.g., the long-term study by \citealt{2009ApJ...697L...1K,2009ApJ...697..592T}). 
For two {\sl Chandra} observations a few months before, 2009 July 31 (observation ID 10053) and August 6 (observation ID 10932; PI Nanda Rea), with orbital phases $\phi=$0.81-0.95 and 0.35-0.43, we found fluxes 8.7$\times10^{-12}$\,erg\,cm$^{-2}$\,s$^{-1}$ and 7.7$\times10^{-12}$\,erg\,cm$^{-2}$\,s$^{-1}$, respectively, close to the values previously measured at these phases. 
 
With a count rate of 0.27 counts per frame, the pile-up fraction would 
be 10\% if the target were on-axis (estimated using {\sl Chandra} PIMMS).
As the source was $\approx 3'$ off-axis in our observation,
the actual pile-up fraction is about 4\%, as we estimated with the aid of MARX simulations (see below).

\begin{deluxetable}{ccccccc}
\tablecaption{Power-law spectral fits\label{fit}}
\tablewidth{0pt}
\tablehead{
\colhead{Region} & \colhead{Radius range (\arcsec)} & \colhead{$N_H$ (cm$^{-2}$)}  & \colhead{$\Gamma$}& \colhead{$A$\tablenotemark{a}} & \colhead{$\chi^2/{\rm dof}$} & \colhead{$F_{\rm 0.5-8\,keV}^{\rm unabs}$\tablenotemark{b}}
}
\startdata 
Point source & 0-5 & $6.4(6)\times10^{21}$ & 1.44(7) & 35.1(3) & 139.3/176 & 25.4(2)\\
Extended & 20-60 & 6.4$\times10^{21}$ \tablenotemark{c} & 1.9(7) & 2.9(7) & 47/40 & 1.4(3)\\
Extended & 60-120 & 6.4$\times10^{21}$ \tablenotemark{c} & 3.1(5) & 8(2) & 70/103 & 2.4(6)
\enddata
\tablecomments{Fits were performed in the 0.5--8\,keV energy range. Numbers in parentheses are $1\sigma$ uncertainties on the last digit. Confidence contours are shown in Figure \ref{contour}.}
\tablenotetext{a}{Normalization in $10^{-5}$\,photons\,cm$^{-2}$s$^{-1}$keV$^{-1}$ at 1\,keV.}
\tablenotetext{b}{Unabsorbed energy flux in $10^{-13}$\,erg\,cm$^{-2}$s$^{-1}$.}
\tablenotetext{c}{Held fixed in the fit.}
\end{deluxetable}

Our best-fit photon index, $\Gamma=1.44\pm0.07$, is  only marginally consistent with those
previously measured at the same orbital phase, $\Gamma=$1.55-1.6 \citep{2009ApJ...697L...1K}.

\begin{figure*}
\begin{center}
\includegraphics[width=\hsize]{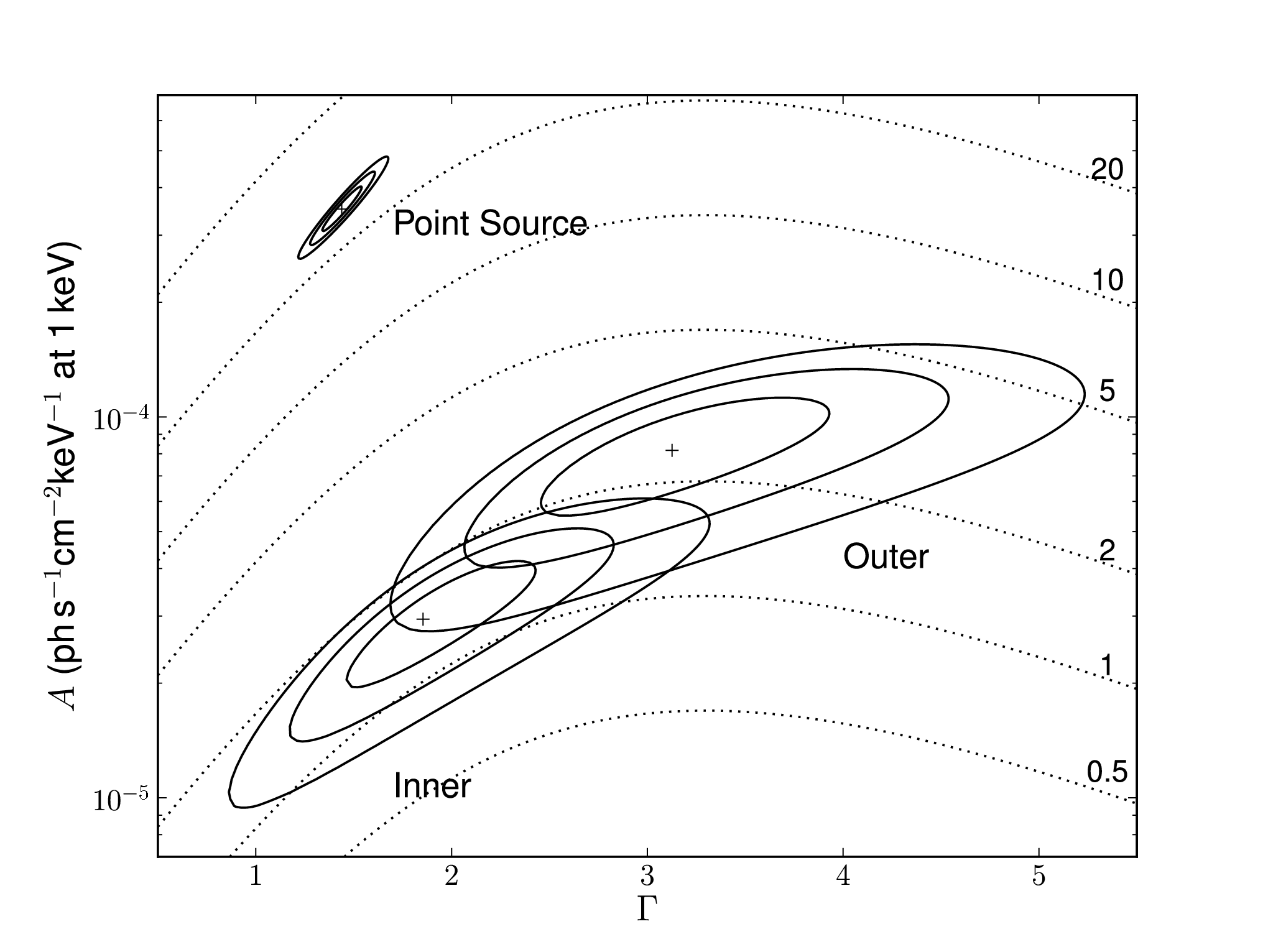}
\caption{Confidence contours at one-, two- and three-sigma, for the PL fit to the spectra. The fits to the annular spectra have $N_H$ fixed to the value obtained for the point source. Lines of constant unabsorbed flux in the 0.5--8\,keV band are shown with dotted lines, and labeled in units of $10^{-13}$\,erg\,s$^{-1}$cm$^{-2}$.
 }\label{contour}
\end{center}
\end{figure*}

\subsection{Extended emission}
 We used {\sl Chandra}\/
Ray Tracer (ChaRT)\footnote{See \url{http://cxc.harvard.edu/chart/threads/index.html}.} and 
MARX\footnote{See \url{http://space.mit.edu/CXC/MARX/}.} software to simulate the PSF for the given spectrum and location on the detector. In order to reduce the Poissonian noise, we made a simulated event file for 200\,ks exposure, filtered the same way as for real data, and then scaled by the exposure times. We used ChaRT with the input spectrum derived from our spectral fit to the central point source and MARX Version 4.4 with parameters {\tt ACIS\_Exposure\_Time=2.2} (to account for the non-standard sub-array mode) and {\tt DitherBlur=0.27} (a measure of the aspect reconstruction accuracy and pixelization by the detector, typically near 0\farcs3 for ACIS). All other parameters were left at their default values.

\begin{center}
\begin{figure*}
\includegraphics[width=0.8\hsize]{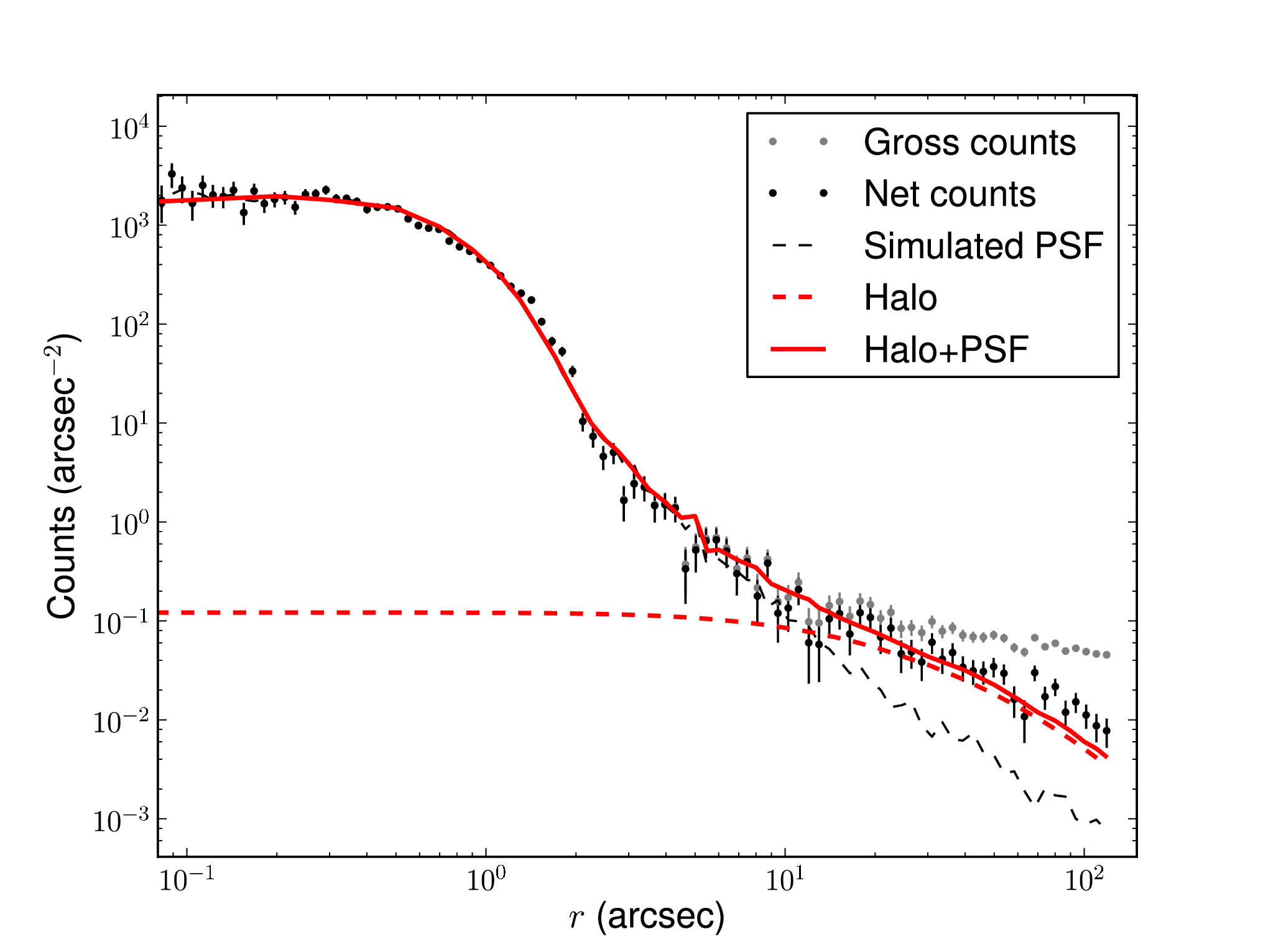}
\includegraphics[width=0.8\hsize]{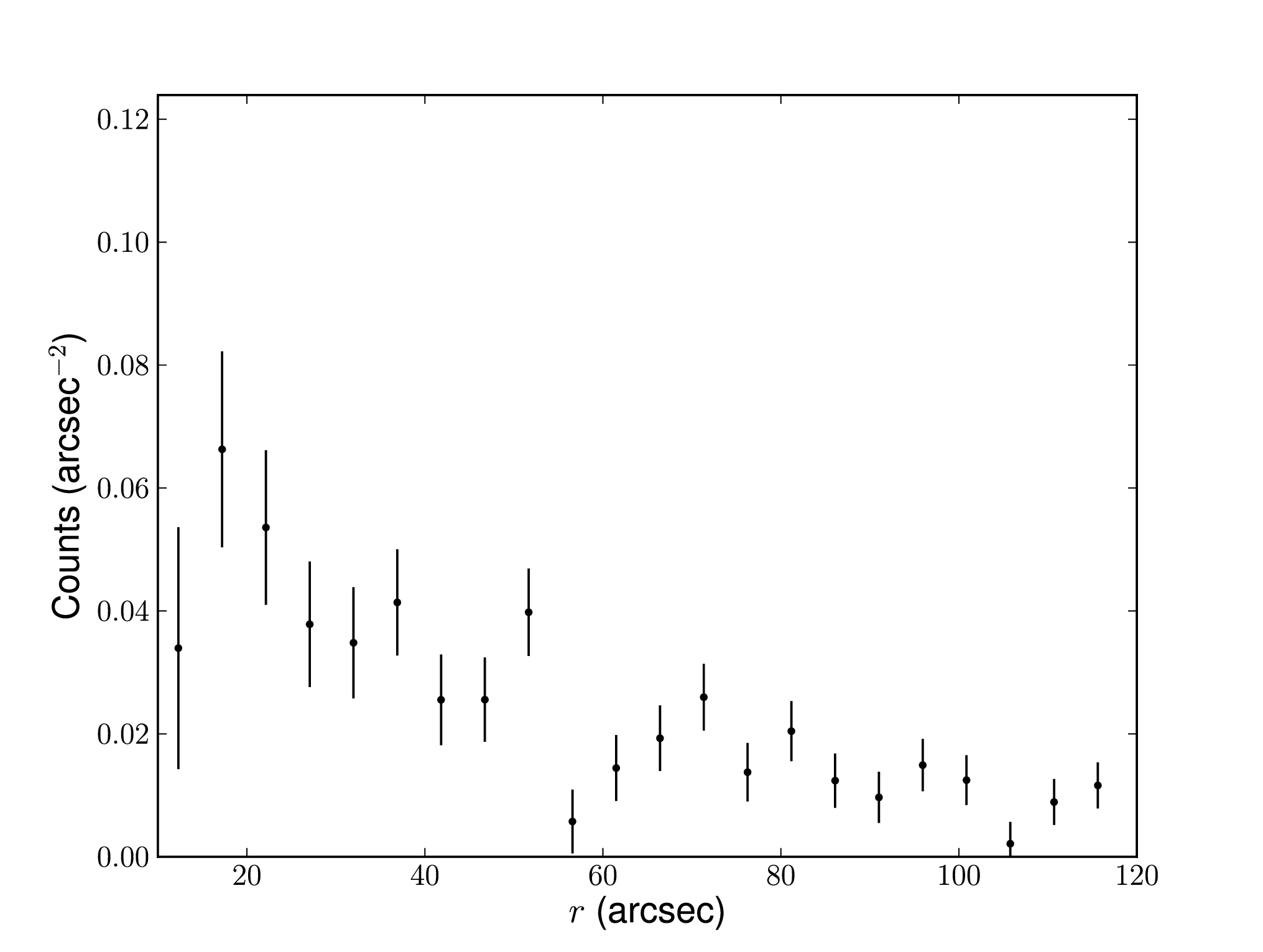}
\caption[Radial plots]{{\it Top}: Radial intensity plot around the point source, for the total (gross), background subtracted (net), simulated PSF count intensities and dust scattering halo model in the energy range 0.59--6.8\,keV. 
{\it Bottom}: The difference between the net and PSF curves, i.e., the excess emission around the point source. }\label{rad}
\end{figure*}
\end{center}

The simulated PSF is shown in Figure \ref{rad}. We calculated the signal-to-noise ratio 
(S/N) of the extended emission near the point source, and found the optimum energy range 
0.59--6.8\,keV. In the radii range 20\arcsec--60\arcsec, we find 745 counts, of which expect 378 from the background and 81 from the PSF wings, giving
$286\pm 29$ source counts, a detection significance of $9.9 \sigma$.
In the radii range 60\arcsec--120\arcsec, there are 1766 counts, of which 1278 should be background and 32 from the PSF wings, giving
$456\pm 51$ source counts, a detection significance of about $8.9\sigma$.

\begin{figure*}
\begin{center}
\includegraphics[width=\hsize]{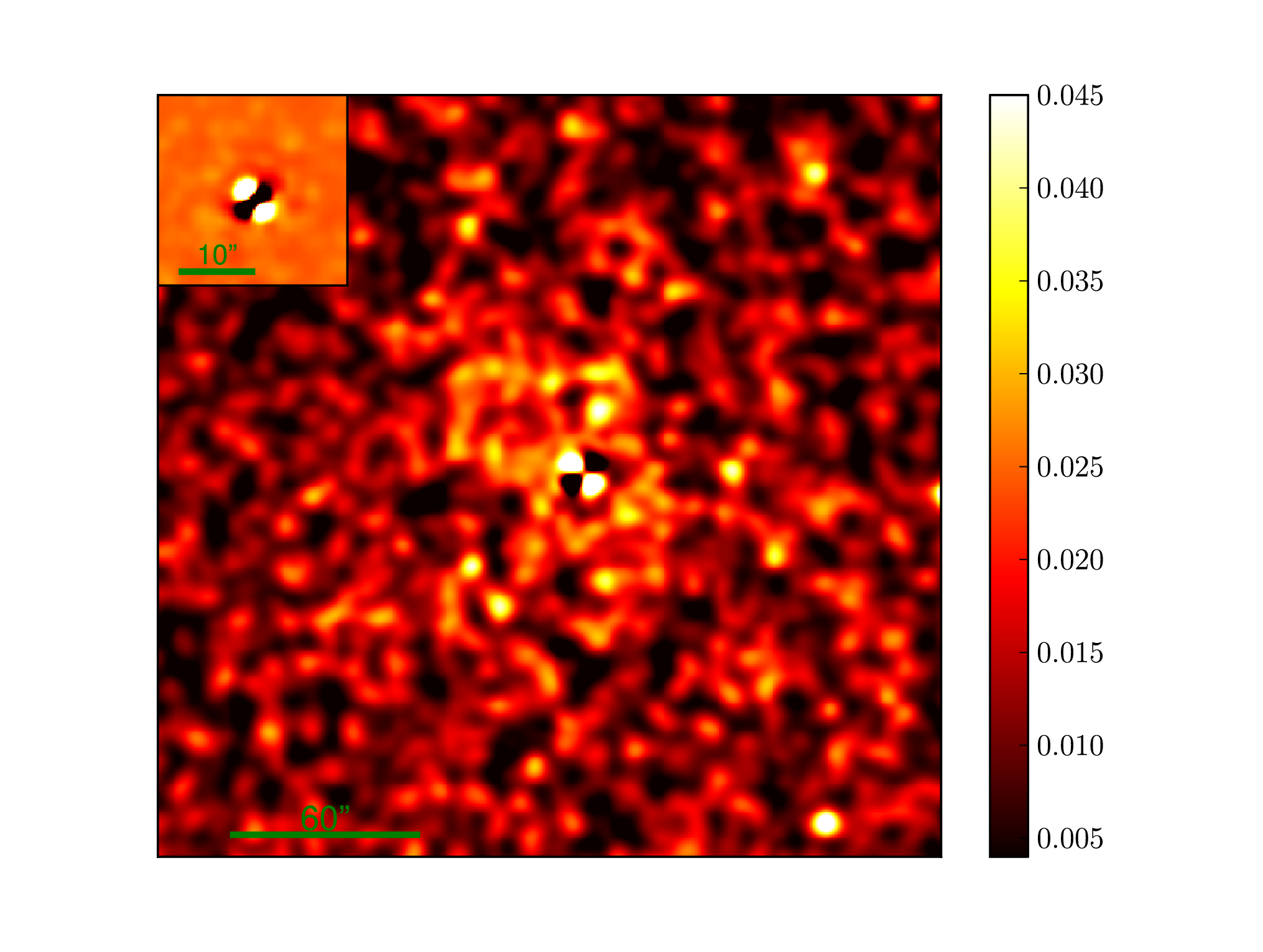}
\caption{256\arcsec$\times$256\arcsec\ image of the vicinity of \ls\ after subtraction of the simulated PSF. The color-bar scale is counts per 0\farcs49$\times$0\farcs49 pixel. The image has been smoothed by a Gaussian kernel with $r=3\arcsec$. The inset shows the central 27\arcsec$\times$27\arcsec\ section, where binning was 0\farcs25, and smoothing width $r=0\farcs5$. It better demonstrates the residual structure around the point source, likely caused by imperfections in the PSF model. }\label{subtract}
\end{center}
\end{figure*}

\begin{figure*}
\begin{center}
\includegraphics[width=\hsize]{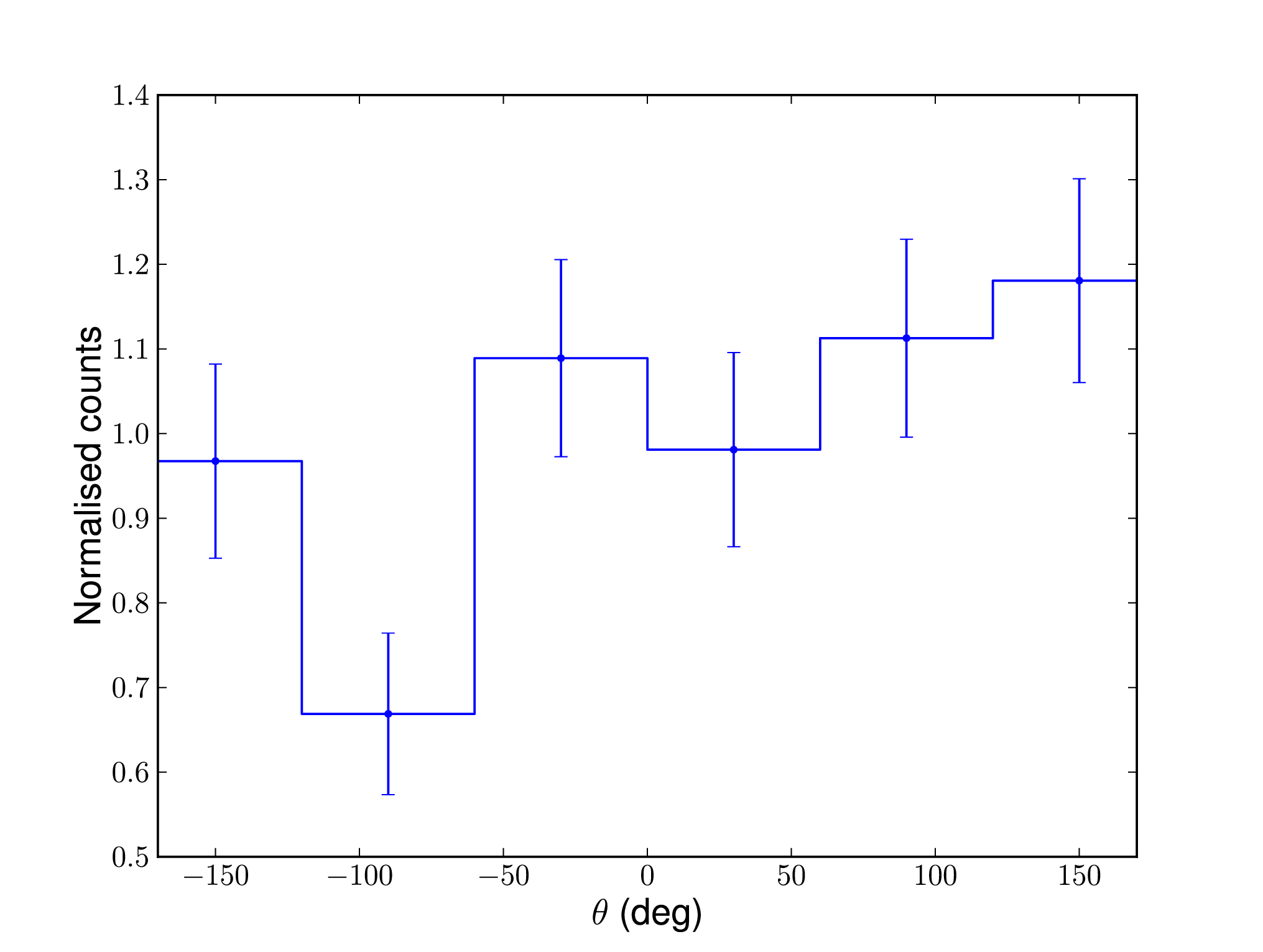}
\caption{Histogram of total counts in the $r=20\arcsec$--60\arcsec  annulus and optimal energy range (0.59--6.8\,keV), as a function of 
azimuthal angle about the point source, with zero angle west and increasing towards the north.}\label{quad}
\end{center}
\end{figure*}

To analyze the spectrum of the extended emission, we extracted counts from the two annular regions, $r=20\arcsec$--60\arcsec and  $r=1\arcmin$--2\arcmin, using the CIAO task {\tt specextract}. The background regions were the same as for the point source above (see Figure \ref{im}). The spectra were once more grouped to at least 25 counts per bin, and fitted with an absorbed PL model in the energy range 0.5--8\,keV. The column density was kept fixed at the best-fit value obtained from the fit to the point source spectrum  ($N_H=6.4\times10^{21}$\,cm$^{-2}$). For the inner annulus, there is considerable contamination from the PSF wings, which may have a different spectrum from the point source because of the energy dependence of the the PSF. To account for this, we fit the counts extracted from the simulated PSF events in the same annulus, and then include this as an additional frozen component ($\Gamma=0.59$, $N=3.83\times10^{-7}$pn\,s$^{-1}$cm$^{-2}$keV$^{-1}$ is 1\,keV) when fitting the extended emission. The results of the fits are shown in Table \ref{fit}, while the corresponding fluxed spectra and confidence contours are shown in Figures \ref{fluxed} and \ref{contour}, respectively. The observed energy flux was $F_{\rm 0.5-8\,keV}=(8.8\pm1.9)\times10^{-14}$\,erg\,cm$^{-2}$s$^{-1}$ (inner) and $(7.5\pm1.9)\times10^{-14}$\,erg\,cm$^{-2}$s$^{-1}$ (outer). 
Using the `optimal' energy range as defined above gives very similar fit parameters. We find that the inner annulus spectrum is only slightly softer than the point source, while the outer annulus is significantly softer.

To better see the morphology of the extended emission, we subtracted the model PSF from the original image. The relative positions of the simulated and observed sources were adjusted to minimize the residuals.
 The result is shown in Figure \ref{subtract}. Some extended structure is apparent, such as wisps to the north and east. With the relatively low 
S/N it is hard to say more, but, interestingly, similar structure 
 is visible in the short on-axis Chandra observation of 2005 April 13 (Figure \ref{im}).
To quantify any anisotropy, we considered the distribution of the extended emission over azimuthal angles about the point source. 
Figure \ref{quad} shows a net histogram of background- and PSF-subtracted counts for the inner annulus. The distribution 
 appears to be non-uniform,
 with $\chi^2=15.9$ for 5 degrees of freedom for a constant value (null hypothesis probability of 0.7\%).

The difference between the simulated and point source images shows  residuals up to $r=3\arcsec$, although the two agree well in the radial plot (Figure \ref{rad}). This suggests some imperfections in the PSF model on small scales.

\subsection{Transient field sources}\label{bob}
A number of small brightness enhancements are apparent in the images in Figure \ref{im}. We have labeled the significant ones with white markers. They are all consistent with being point-like, although with the small numbers of counts, this is not very restrictive.

\begin{figure*}
\begin{center}
\includegraphics[width=0.8\hsize]{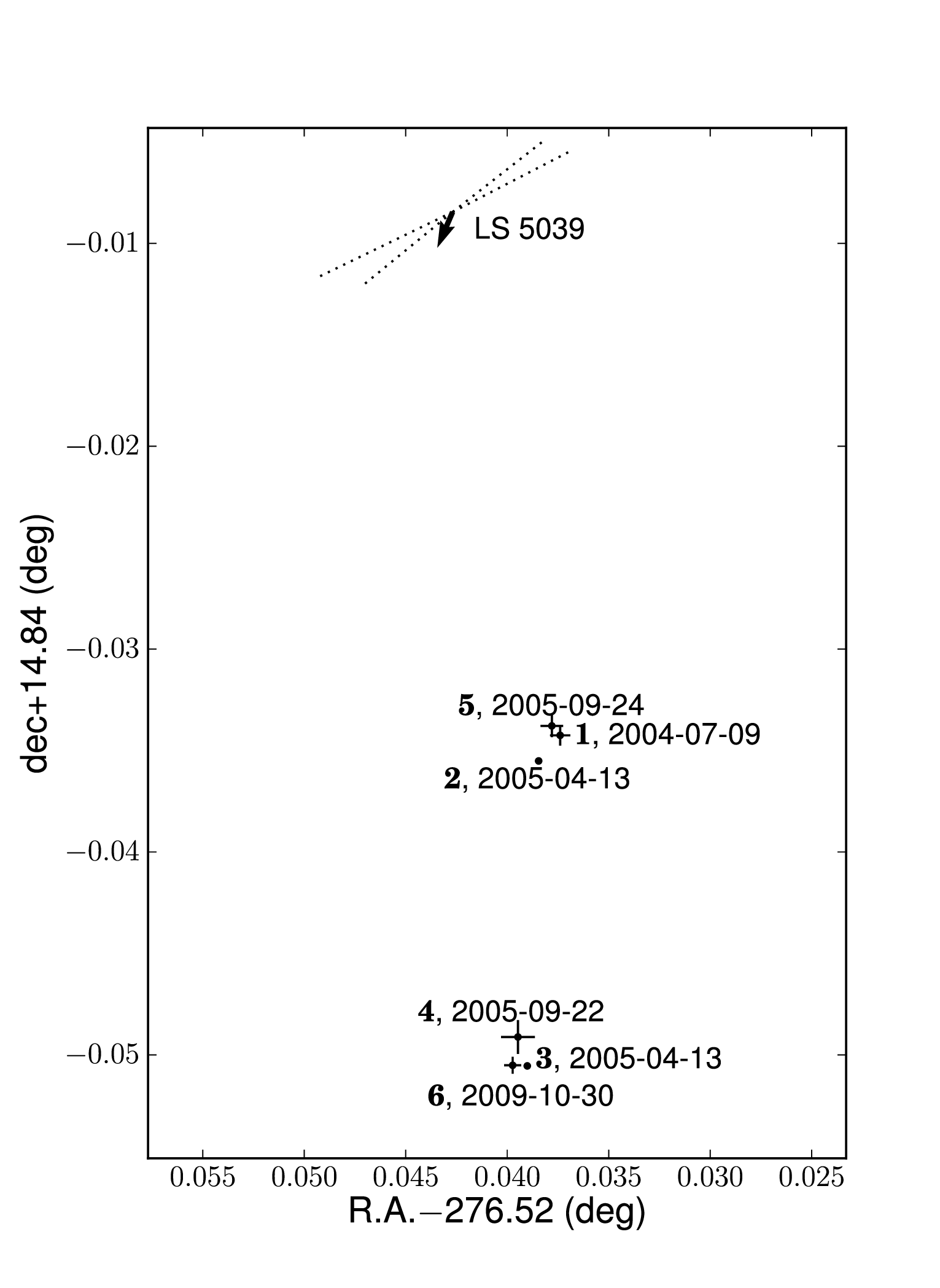}
\caption{Detected sources in the field of \ls. We show \ls's proper motion vector (1000 times yearly value measured by \citealt{2002A&A...384..954R},
black arrow) and the position angles of radio elongation at two different epochs (dashed lines; \citealt{2008A&A...481...17R}).
The object numbers (bold) refer to the white labels in Figure \ref{im}.}\label{blob}
\end{center}
\end{figure*}

\begin{deluxetable}{cccc}
\tablecaption{Fluxes of transient field source fluxes\label{blob_flux}}
\tablewidth{0in}
\tablehead{
\colhead{Date} & \colhead{Sources} & \multicolumn{2}{c}{ Flux\tablenotemark{a} } \\
 && \colhead{North} & \colhead{South}
} 
\startdata 
2004-07-09 & 1&$3.6\pm0.8$& $<$4.4\\
2004-07-11 & &$<$0.8& $<1.4$\\
2005-04-13 & 2,3&$2.4\pm1.1$& $2.3\pm1.7$ \\
2005-09-22 & 4&$<$0.7& $1.3\pm0.3$\\
2005-09-24 & 5&$3.1\pm0.5$& $<$2.0 \\
2009-10-30 & 6&$<$0.47& $1.6\pm0.4$ \\
\enddata
\tablenotetext{a}{Observed, background-subtracted energy flux (1.0--6.8\,keV) in units of $10^{-14}$\,erg\,cm$^{-2}$s$^{-1}$. Uncertainties are at 1-$\sigma$ limits are at 3-$\sigma$. Columns `North' and `South' represent sources in the two apparent spatial groupings, see Figures \ref{blob} and \ref{app}.}
\end{deluxetable}

The numbered sources in Figure \ref{im} are projected onto the same sky coordinates in Figure \ref{blob}. They appear to lie within a narrow stripe almost due south from \ls\ in two clumps or clusters. All the sources have  rather hard spectra (mean energies 2--4\,keV). Notably, the source that was bright on 2004 July 9 
(Source 1) had completely disappeared two days later (even though the latter observation had a longer exposure time, and would  have detected it); similarly Source 5 appeared on a similar time-scale. We cannot exclude that sources 1 and 5 are the same object, see below. The fluxes of these sources are given in Table \ref{blob_flux}.

\begin{figure}[h]
\includegraphics[width=0.45\hsize]{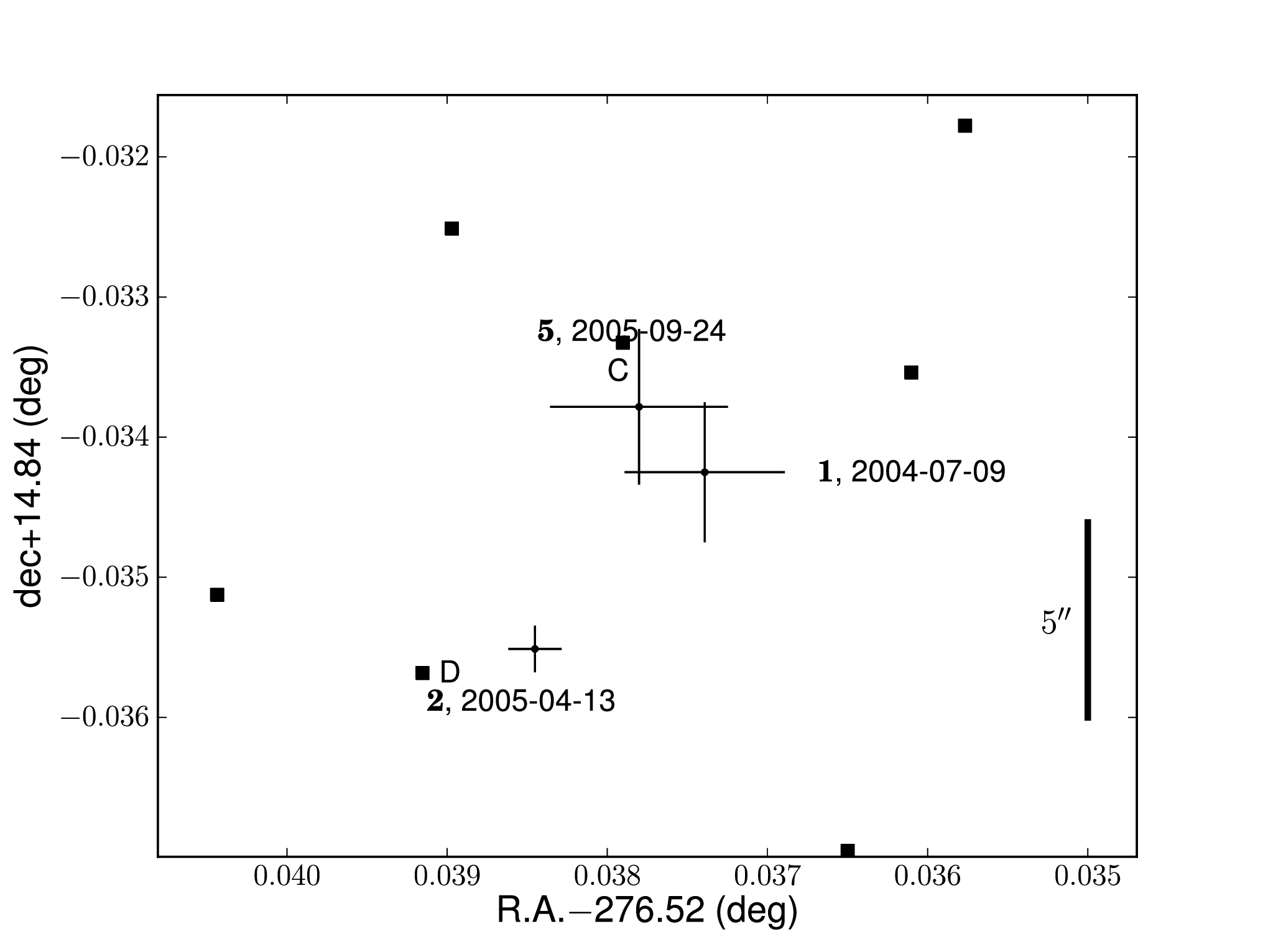}
\includegraphics[width=0.45\hsize]{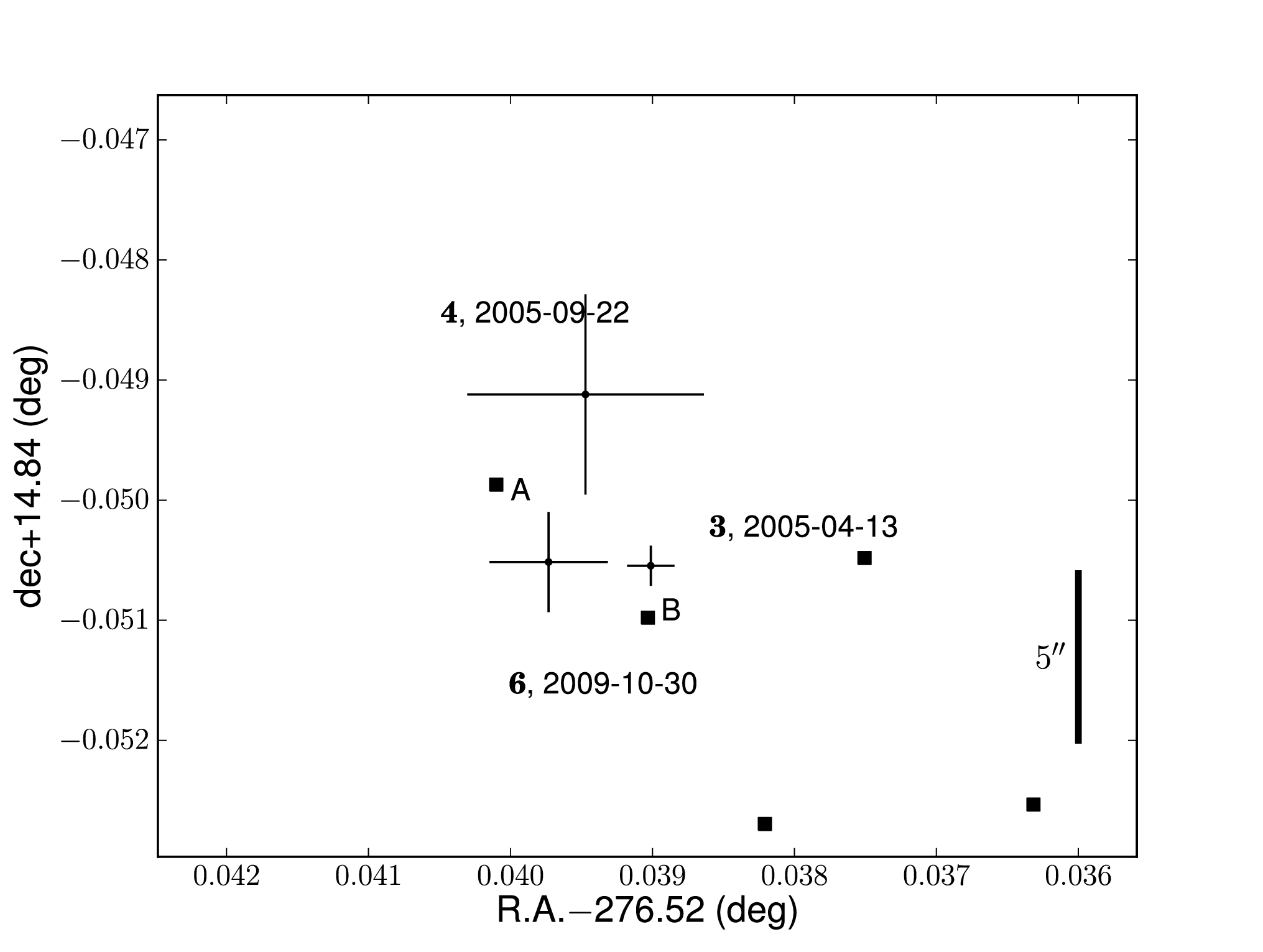}
\caption{Close-up of the locations of field sources shown in Figure 7, northern group (left) and southern group (right). Also shown are sources from the 2MASS catalog (squares). The nearest stars are labeled by letters, and their photometric data and positions are given in Table \ref{app_phot}. The coordinates shown here are after the application of a bore-sight correction. Scale bars are 5\arcsec\ in length.}\label{app}
\end{figure}

We show zoomed-in views of the locations of the field sources detected during observations of LS 5039. In Figure \ref{app} the positions are plotted, after bore-sight correction, along with sources from the 2MASS catalog \citep{2006AJ....131.1163S}. The nearest 2MASS sources, which have positions consistent with (some of) the field sources, are listed in Table \ref{app_phot}  their infra-red magnitudes. Where possible, we also include optical magnitudes. 

Sources 2 and 3 are only marginal detections. If they are not real sources, or if they are unrelated to the rest of the sources detected, then it is likely that sources 1 and 5 are a single flaring source  possibly associated with 2MASS source C. Likewise, sources 4 and 6 could be a single persistent source, perhaps associated with 2MASS source A. 

\begin{deluxetable}{cccccccc}
\tablecaption{Photometry and positions of 2MASS sources near faint X-ray sources \label{app_phot}}
\tablewidth{0in}
\tablehead{
\colhead{Label} & \colhead{2MASS ID} & \colhead{$J$} & \colhead{$H$} & \colhead{$K_{\rm S}$}  & \colhead{Other} & \colhead{R.A. (\degr)} & \colhead{decl (\degr)} \\
& & & & & \colhead{photometry\tablenotemark{a}}
}
\startdata 
A & 18261441-1453235 & 14.40 & 12.62& 11.91 & &276.560075 & $-$14.889871\\
B & 18261416-1453275 & 13.68 & 12.64& 12.22 & &276.559035 & $-$14.890979\\
C & 18261389-1452239 & 14.40 & $>$13.50& $>$13.28 & $B=16.64,V=16.20$ & 276.557879 & $-$14.873326\\
D & 18261419-1452324 & 12.50 & 11.03& 10.46& $R=19.4$ & 276.559134 & $-$14.875684\\
\enddata
\tablecomments{~Labels refer to the those shown in Figure \ref{app}. 2MASS photometric uncertainties are 0.04-0.06\,mag.}
\tablenotetext{a}{Magnitudes noted in the NOMAD catalog.}
\end{deluxetable}

\section{Discussion and conclusions}\label{dis}

 Our analysis has revealed a highly significant extended emission discernible  up to $\simeq2'$  from LS 5039 (limited by the FOV boundaries). This emission could be either due to the  dust scattering of X-rays emitted by the unresolved source or it could be emission from relativistic particles escaping from the vicinity  of  LS 5039.  Below we discuss both options.    In addition, we observed at least two faint unresolved  sources.  These could be field stars, background highly-absorbed CVs or quiescent XRBs (or even AGNs seen through the Galactic plane), but they might in principle  be manifestations of a  collimated  outflow from   LS 5039.
      
The flux of the point source LS 5039 was a factor of $\sim$2 lower than expected for its orbital phase, despite previously appearing to have a stable orbital light-curve \citep{2009ApJ...697L...1K}. Previous studies did suggest larger variability \citep{2003A&A...405..285R,1999A&A...347..518R}, but the use of different instrumentation (particularly RXTE) made the variability hard to verify. A decrease in flux could perhaps be explained by a change in accretion state (if the compact source is a BH) but even for  pulsar emission, it can be explained by  changes in the wind properties of the companion
O-star which also manifest themselves via the variability of H$_{\alpha}$ emission \citep{2003A&A...405..285R}.

\subsection{Dust scattering halo}

Scattering of X-rays emitted by the unresolved source off the dust lying along the line of sight produces extended emission known as a {\em dust scattering halo} (e.g., \citealt{1995A&A...293..889P}).  The halo brightness scales linearly with the intervening dust column, which is usually assumed to be proportional to the hydrogen column density, $N_H$, measured from  X-ray spectra, i.e.,  scattering optical depth at 1\,keV 
is $\tau_{{\rm s,1keV}}\simeq S (N_H/10^{22}~{\rm cm}^{-2})$,
 where  the proportionality constant $S$ can only be determined empirically and depends on properties of the intervening dust which may differ for different objects.   
 For instance,  \citet{1995A&A...293..889P}   found  a mean 
$S\simeq 0.5$  from their sample of X-ray halos, which 
 corresponds  $\tau_{{\rm s,1keV}}\simeq0.24$ for LS 5039 where $N_H=6.4\times10^{21}$ cm$^{-2}$ (Table 2). 
An even better correlation 
for $\tau_{{\rm s}}$ was obtained with
 the optical extinction towards the object,   $\tau_{{\rm s,1keV}}=0.056 A_V$  (see Appendix A), which gives $\tau_{{\rm s,1keV}}=0.22$ (for $A_V=3.9$; \citealt{2002A&A...384..954R}), in  good agreement with the previous estimate.  The relatively low value of $\tau_{{\rm s,1keV}}$ allows us to neglect the contributions of multiple scatterings for $E\ga1$\,keV. 

We have calculated the halo profile in the single-scattering approximation  by folding the spectral intensity of a halo (see e.g., \citealt{1991ApJ...376..490M}, also Appendix A) with the detector response in the 0.5--8\,keV energy range.  We find that, for instance,
 for the parameters $\Theta=360\arcsec$ and $S=1$, and the dust distribution  function $f(x)$  defined in Appendix A,
the dust halo model overall describes the observed  radial profile (see Fig. \ref{rad}). Although some deviations are noticeable, they might be attributed to the imperfect choice of the above parameters (we did not perform rigorous fitting) or to the inaccuracy  of the \citet{2003ApJ...598.1026D} model or the Rayleigh-Gans (RG) approximation to the scattering cross section at low energies ($\lesssim 2$ keV). 

The dust distribution function $f(x)$ used to calculate the halo model (see
Equation (A8) and Figure \ref{rad})
assumes a lack of absorption/scattering in the immediate vicinity of the
source.
 Such dust distribution 
is consistent with the lack of evidence of significant intrinsic absorption in the \ls\ binary.   The upper
limit on the intrinsic column density is as low as  $2.6\times10^{21}$\,cm$^{-2}$ \citep{2007A&A...473..545B,2009ApJ...697..592T},
while the total absorbing column density measured from X-ray spectra changes by  $\lesssim 10\%$ \citep{2009ApJ...697..592T} with binary phase.
Low intrinsic absorption would be expected if most of  X-rays are emitted far from the donor star, where its wind is sufficiently rarefied   (e.g. \citealt{2011MNRAS.411..193S}).  

We should also point out that,  in addition to the dust distribution profile, the   dust scattering model we use has two free parameters ($S$ and $\Theta$) which can attain 
 values within rather broad ranges, depending on the actual properties of the dust grains. Therefore, the mere fact that the model qualitatively fits the observed brightness 
 distribution does not guarantee that the extended emission is a dust halo. In fact,  the observed  azimuthal  asymmetry and the hard spectrum of the inner extended emission would be difficult to explain by a dust halo.  Significant  spectral softening is expected from halo models. Specifically, for the dust model used, the spatially integrated (between $r=20''$ and $60''$) spectrum  is expected to have $\Gamma_{\rm halo}\simeq \Gamma_{\rm point}+2\simeq3.5$, because the scattering cross-section $\sigma_s(E)\propto E^{-2}$ (see Appendix A). Instead, we find a significantly harder spectrum, $\Gamma\approx1.9$. On the other hand, the outer extended emission is indeed much softer and more symmetric, hence could have some contribution from a halo.


\subsection{Extended nebula of LS 5039}

 The nature of the compact object (BH or NS) in \ls\ remains  subject of a debate.  \ls\  (and \lsi) are quite different in their temporal and spectral properties from 
  other $\mu$QSOs and HMXBs, which show transitions  between different states and large variations in luminosity. These two systems are also, to date, the only $\mu$QSOs whose VHE emission has been firmly detected\footnote{The only other HMXB firmly detected at VHE is B1259$-$63, where a young pulsar orbits a Be star on a 3.4\,yr, highly eccentric orbit.}.   The X-ray light curve of \ls, which  shows remarkable  long-term stability \citep{2009ApJ...697L...1K,2009ApJ...697..592T,2009A&A...494L..37H}, and peculiar, variable radio morphology \citep{2008A&A...481...17R} argue against the accretion scenario.  

If, despite the lack of the usual manifestations of accretion, the compact object in \ls\ is a BH accreting in an unusual regime \citep{2005MNRAS.364..899C},  it may still be possible that it produces relativistic particles, e.g., via the Blandford-Znajek process \citep{1977MNRAS.179..433B} or in an MHD jet.  In this case, an extended nebula could still be formed. 
For the case of a black hole, the majority of high-energy emission is expected to be produced in a jet-type axial outflow (e.g., \citealt{2006A&A...451..259P}). The possible observational manifestations of such outflows and a discussion of their energetics is discussed by \citet{2010arXiv1001.1244R} and their extended emission by \citet{2008ApJ...686.1145H}. The relativistic electrons responsible for the $\gamma$-ray emission in the inner jet will cool with distance. Flow speeds are typically close to $c$ \citep{2006smqw.confE.101Y}, and so X-ray emission can be expected out to appreciable distances from the black hole, e.g., the plasma clumps seen by \citet{2002Sci...298..196C} up to $\sim30$\arcsec\ from $\mu$QSO XTE J1550-564. Furthermore, the spin axis of the black hole may precess, and so the orientation of the outflow change with time, as with SS 334 \citep{2007A&A...474..903B} and thus fill an extended volume with energetic emission, rather than the more typical collimated outflows. Such large-angle precession may be hard to explain, however, possibly requiring a third massive component to the system.

A possible alternative interpretation of the high-energy emission from \ls\  is a pulsar wind nebula (PWN) powered  by a young, energetic pulsar. In this scenario, modeling of the intra-binary shock  suggests that the O-star wind dominates dynamically over the pulsar wind, which makes  the shocked wind  flow away from the 
head of the intra-binary bow-shock and
along the   bow-shock surface which is curved away from the O-star (see e.g., Figures 1 and 2 in \citealt{2011MNRAS.411..193S}).  If the opening angle of the bow-shock is small, it would resemble a cometary tail rotating around the O-star together with the pulsar and pointing in the direction of the O-star wind (cf.\  variable radio morphology of  \lsi\ resolved with VLBI; \citealt{2006smqw.confE..52D}; see also the discussion in \citealt{2011ApJ...730....2P}). Emission from the innermost regions of the bow-shock may be responsible for  the unresolved, bright X-ray source whose surface brightness varies with the orbital phase owing  to Doppler boosting 
 \citep{2010A&A...516A..18D}.  Since the tail is rotating, it winds itself into a tight spiral, which at larger distances  from the pulsar and with limited angular resolution would appear  as disk-like emission concentrated toward the orbital plane.
   Moreover,  it   seems reasonable to assume that, in this disk, the pulsar wind becomes well mixed with the dynamically dominant companion wind and that both are advected away at the O-star wind terminal speed, $v_{\infty}$,
 at distances $r\gg a$ (where $a$ is the orbital separation).  Adopting  $v_{\infty}=2400$\,km\,s$^{-1}$ 
\citep{2004ApJ...600..927M},
  we can estimate the time it takes for the wind to reach the angular distance $\alpha$ as t$_{\rm dyn}\simeq300 (\alpha/1\arcmin)(v_{\infty}/2400~{\rm km~s}^{-1})^{-1}$\,yr.

 The likely  emission mechanism in this case would be synchrotron radiation (see below), implying the cooling time   $t_{\rm syn}\sim300 (E/5~{\rm keV })^{-1/2}(B/10~\mu{\rm
    G})^{-3/2}$ yrs.  Requiring $t_{\rm syn}\gtrsim t_{\rm dyn}$ places an upper limit on the average magnetic field in the extended emission region, \\ $B\lesssim10  (E/5~{\rm keV })^{-1/3}(v_{\infty}/2400~{\rm km~s}^{-1})^{2/3}(\alpha/1')^{-2/3}$\,$\mu$G, where $E$ is the observed energy of synchrotron photons  emitted by relativistic electron with Lorentz factor \\
    $\gamma\sim3\times10^8 (E/5~{\rm keV })^{1/2}(B/10\,\mu{\rm  G})^{-1/2}$. 
    Since the upper limit on $B$ is not much higher than  the value of the magnetic field in the ISM,  it is likely that the actual magnetic field is on the order of 10 $\mu$G,
 and hence the 
     electron Lorentz factor should be of the same order as those in ordinary PWNe around isolated pulsars.  On the other hand, the inferred $\gamma$  is close to the maximum possible Lorentz factor 
estimated by \citet{2006A&A...456..801D}  by requiring  the electron gyro-radius  to be less than the size of the acceleration region,  assumed to be on the order of the standoff distance  for the intra-binary shock. 
Moreover,  \citet{2006A&A...456..801D}  
concluded that electrons will not attain $\gamma\gtrsim10^{8}$ 
because of the strong synchrotron radiative losses in the innermost regions. Additional acceleration outside the intra-binary shock  (e.g., via magnetic reconnection in the spiral-like tail) may provide a  way to circumvent this limitation. Alternatively, 
$\gamma$ could be somewhat lower if  the magnetic field is higher than estimated above, which is possible if the bulk flow speed if somewhat higher than the terminal O-star wind speed. Indeed, up to a certain  distance, the fast pulsar wind could provide some re-acceleration to the O-star wind until both components become well mixed. However, the detailed modeling of this  complex interaction is beyond the scope of this paper.   The observed softening of the extended emission with the distance from the point source could be attributed to  synchrotron cooling which can be substantial at $\alpha\gtrsim1'$  if  $B\sim 10$\,$\mu$G or somewhat larger.

 Inverse Compton (IC) and adiabatic expansion  cooling  compete with synchrotron cooling  and can even dominate, depending on the distance from the particle acceleration region (the intra-binary shock).
 Although  at the distances 
 of interest here, the O-star radiation energy density\footnote{Following 
\citet{2011MNRAS.411..193S},  we assume  $R_{\star}=9.3R_{\odot}$ and $T_{\star}=39,000$ K for the O-star. }, $\epsilon_{\rm rad}=3.8\times10^{-10}(\alpha/1')^{-2}$ erg cm$^{-3}$,  would exceed the magnetic field energy density, $\epsilon_{\rm B}=4.0\times10^{-12}(B/10~\mu{\rm
    G})^2$ erg cm$^{-3}$, the IC cooling of  $\gamma\gtrsim6\times10^4$ electrons occurs in the Klein-Nishina regime and is hence inefficient in comparison with the  competing processes (see \citealt{1970RvMP...42..237B}).  Indeed, the relevant IC cooling time, $\tau_{\rm IC}\sim 20 (E/5~{\rm keV })^{1/2}(B/10\,\mu{\rm
    G})^{-1/2} (\alpha/1')^{2}$\,Myr, on the O-star light  will be much larger than the corresponding synchrotron cooling time of electrons emitting $E=5$\,keV photons in $B=10$~$\mu$G.  On the other hand, the same electrons will also cool via IC upscattering of CMB photons, and this process will dominate over the IC scattering on O-star light at the distances of interest. 
    IC scattering off the CMB proceeds in Thompson regime with characteristic cooling time $\tau_{\rm IC,CMB}\simeq 7 (E/5~{\rm keV })^{-1/2}(B/10\,\mu{\rm
    G})^{1/2}$\,kyr, which is still larger than the corresponding $\tau_{\rm syn}$ (see above).   Finally,  there is also cooling due to adiabatic expansion,
 which, in 2D geometry,      has a characteristic  cooling timescale  comparable to dynamic timescale     (see, e.g. Eqn.\ (28) in \citealt{2006A&A...456..801D}).  

 \citet{2010MNRAS.403.1873Z} suggested that the X-ray emission from the compact (unresolved) PWN in \lsi\ could be  produced via IC scattering of stellar photons by the pulsar wind electrons with $\gamma\sim 20$.  The synchrotron emission by the same electrons is expected to be in the radio, while the GeV emission could be interpreted as IC up-scattering of stellar light by more energetic photons with $\gamma$ up to $2\times10^4$. If the electron SED in \ls\   extends down to such low $\gamma$,  these processes may contribute significantly at sufficiently small distances from the binary. As the  low-energy electrons  should remain unaffected either by IC or synchrotron cooling on any plausible dynamical timescales, one would expect to see radio synchrotron emission on much larger angular scales than it has been reported. Deep radio observations,  sensitive to extended structures on arcsecond--arcminute scales, can provide a useful diagnostics in this case.

From modeling of absorption and occultation,
 \citet{2011MNRAS.411..193S} estimated the plausible range of $\dot{E}=(0.1$--$38)\times10^{36}$\,erg\,s$^{-1}$
for the alleged pulsar in \ls. The observed unabsorbed 
luminosity\footnote{Here we take into account only the inner extended emission, because the outer region may include a significant dust scattering contribution.} $L_{\rm 0.5-8\, keV}=1\times10^{32} d_{2.5}^2$\,erg\,s$^{-1}$  implies the radiative efficiency range $3\times10^{-6}$--$10^{-3}$, which well matches 
the range for resolved PWNe around subsonically-moving isolated Vela-like pulsars  ($10^{-5}$--$10^{-3}$; Figure 7 in \citealt {2007ApJ...660.1413K,2008AIPC..983..171K}).
 If \ls\ is indeed associated with  SNR G16.8$-$1.1, it has a kinematic age of 40--150\,kyr, i.e.,  a factor of a few older than the Vela-like pulsars listed in Table 2 of   \citet{2007ApJ...660.1413K}. This would favor lower values of efficiency from the range given above, and, correspondingly,  lower O-star mass loss rates from the range discussed in \citet{2011MNRAS.411..193S}.

If the extended emission is due to the pulsar wind, there can be non-variable TeV emission associated with the extended wind-filled region, even larger than that resolved in X-rays, resembling relic PWNe around isolated pulsars (see e.g., \citealt{2010AIPC.1248...25K}).  Indeed, \citet{2006A&A...460..743A} reported orbital modulation of the LS 5039 TeV flux peaking near inferior conjunction phase but also reported a second component with a different spectrum seen near the light curve minimum. While the harder emission at the peak of the light curve is likely to come from the innermost region of the binary, the soft spectrum of the second component is akin to those of TeV PWNe (see \citealt{2010AIPC.1248...25K} and references therein) and could  be attributed to the extended PWN.  Therefore, further high-sensitivity, high-resolution observations in TeV are certainly warranted.

\subsection{Transient sources in the field}

We have found  several (at least two, see \S3.3)
 transient X-ray sources, all of which are distributed within a very narrow strip south of \ls.  Taken together, the positions of these sources at different epochs are inconsistent  with a constant velocity or uniformly decelerated motion of a single source (see Figures 1 and 7). Therefore, they  cannot  represent the propagation of a single clump of plasma ejected by  \ls.
The  sources could be unrelated flaring objects (e.g., heavily absorbed background  CVs in the central region of the Galaxy) or, alternatively, all  could be  flare-ups in an outflow largely invisible in X-rays, in which case they don't have to be causally connected. Although relativistically outflowing plasma clumps have been seen before in $\mu$QSOs, their variability at such large distances from the central source has never been seen as dramatic as would appear to be the case here (e.g., \citealt{2007A&A...474..903B,2002Sci...298..196C}). The direction of this outflow would be different from that of the two radio jets seen on milli-arcsecond scales, but  the latter were seen to change orientation by 12\degr$\pm3\degr$ \citep{2008A&A...481...17R}.  

We have searched for other variable sources within the {\sl XMM-Newton} EPIC-PN FOVs (the smallest FOVs among the  observations; see Figure 1) and did not find any other  variable sources. Further monitoring of LS 5039 with {\sl Chandra} or {\sl XMM-Newton} is needed to draw any definitive conclusions on the nature of the transient sources we discovered. Confirming  ejecta clumps would favor accretion/ejection scenario over the pulsar wind one.

\subsection{Comparison with \lsi\ and B1259$-$63}

Among known HMXBs, the one that resembles \ls\ the most\footnote{A new HMXB, 1FGL J1018.6$-$5856, has been discovered very recently \citep{2010ATel.3221....1C} whose properties are even closer to those of LS 5039 \citep{2010ATel.3228....1G}, but its has not yet been investigated in such detail as \lsi.}  is  \lsi.    In both systems, the donor stars are massive early type stars (O and Be stars with masses 23$M_{\odot}$  and 12$M_{\odot}$, in \ls\ and \lsi, respectively). Both exhibit  very similar X-ray properties (including fluxes, spectral slopes, $N_H$; e.g., \citealt{2010MNRAS.405.2206R}) and are located at similar distances.  
Recently, both systems were detected in  GeV and TeV, where they also share many common properties such as  similar VHE spectra and light curves (\citealt{2010arXiv1008.4762H}, and references therein). Among the differences are  a factor 
 of 7 larger orbital period ($P_{\rm orb}=26.5$ days) and a factor of 2 larger eccentricity ($e=0.72$) of the 
  \lsi\ orbit. Also, \lsi\ appears to be more variable in X-rays, sometimes showing   short flares  \citep{2010ApJ...719L.104T}. This could be due to non-uniformity of the companion wind which is believed to be   concentrated in a rather thin disk in the  equatorial plane \citep{2009ApJ...693.1462S}. The compact object interacts with  this dense wind only during a relatively short phase interval  while it crosses the disk \citep{2009ApJ...693.1462S}. The wind of the O-star in LS 5039 is believed to be much more isotropic. Analyzing a 49\,ks {\sl Chandra}
 ACIS observation of  \lsi, \citet{2007ApJ...664L..39P}  found  faint ($58\pm18$ counts) asymmetric excess emission  extending $\approx12\arcsec$ from the 
  point source. This emission was not confirmed in the subsequent longer (96\,ks) {\sl Chandra} ACIS observation, but that observation used  Continuous Clocking mode with much higher background and only one spatial dimension, greatly complicating the analysis \citep{2010MNRAS.405.2206R}. Finding a much fainter extended emission around \lsi\  while having the point source X-ray properties  very similar to those of the \ls\ point source, supports the conclusion that  most of the extended emission in \ls\  {\em is not a dust scattering halo}, if the gas-to-dust  ratios and dust grain properties are similar for these two systems. If the extended emission around the two sources were both dust halos, then, with a similar dust grain size distribution, the dust-to-gas ratio would need to be a factor of 10 larger towards \ls\ in order to explain the brighter halo. The dust-to-gas ratio is known to have large scatter along different sight-lines, however (see Appendix A).
 
    If the extended emission around \ls\ is due to particles supplied by the pulsar wind, one could speculate that this emission should be fainter in \lsi\ because there the pulsar spends most of the time  outside the dense companion wind,
 and hence the pulsar wind is not as efficiently confined and decelerated as in \ls. This means that in \lsi\ most of the pulsar wind is leaving the binary with large bulk velocity, which should result  in much larger but fainter X-ray nebula, undetectable in existing data.
    
Another object of possibly similar nature is B1259$-$63, a HMXB with a much wider orbit 
($P_{\rm orb}=3.4$\,yr), high eccentricity ($e=0.87$) and a known pulsar
($P=48$ ms, $\dot{E}=8.3\times 10^{35}$ erg s$^{-1}$), where
\citet{2011ApJ...730....2P} found evidence of extended emission. In that case, the emission is confined to $r<15\arcsec$.   This shows that a pulsar system is indeed capable of creating extra-binary extended X-ray emission, and therefore that such a scenario may be valid for the case of \ls, although in this case on a much larger spatial scale.

   \section{Summary}\label{sum}

We have discovered asymmetric 
emission around \ls,  extending up to 2\arcmin\ from the point source. Although it is possible that some of this emission is due to a dust scattering halo,  most of it is likely produced by energetic particles emanating from \ls. Although large-scale energetic jets have been observed emanating from $\mu$QSOs, there is a lack of an obvious mechanism to fill a large extended volume with X-ray-emitting particles; the pulsar scenario for the compact object appears to be preferable. If the wind nebula interpretation of the extended emission is true, one would expect to find an even larger radio and/or GeV/TeV nebula, in sensitive enough observations. We find  transient sources in the field in a narrow stripe south of \ls\ whose nature is unclear.

\acknowledgements{
This work is based on data acquired with the {\sl Chandra}
and {\sl XMM-Newton} X-ray Observatories. 
It was supported by the ACIS Instrument Team contract SV4-74018 (PI G.\ Garmire) and partly supported by archival Chandra grant AR8-9009X, NASA grants NNX09AC84G and NNX09AC81G and the National Science Foundation under grants No.\ 0908733 and 0908611.  The work of GGP was also partly supported by the Ministry of Education and Science of Russian Federation (Contract No.\ 11.G34.31.0001).

We thank the anonymous referee for helpful suggestions.
} 

\clearpage
\bibliography{database}

\appendix
\section{Dust halo model}
Dust halos, often seen around bright point-like X-ray objects, are formed by scattering
of source X-ray photons on dust grains. Here we will only discuss the case of
dust optically thin with respect to the
photon scattering, $\tau_{\rm scat}\lesssim 1$, and consider only
azimuthally symmetric halos
(which implies that the dust distribution across the line of sight (LOS) is uniform within the interval of angles $\theta$ at which we see the halo). In this case
the spectral halo intensity (ph\,cm$^{-2}$ s$^{-1}$ keV$^{-1}$ arcmin$^{-2}$) is given by the equation
\begin{equation}
I_{\rm halo}(\theta, E) = F(E) N_H \int_0^1 dx\, \frac{f(x)}{x^2}\, \frac{d\sigma_s(E,\theta_s)}{d\Omega_s}\,,
\end{equation}
where $F(E)$ is the point source spectral flux (photons\,cm$^{-2}$ s$^{-1}$ keV$^{-1}$), $x=(D-d)/D$ is the dimensionless distance from the X-ray source to the
scatterer ($D$ and $d$ are the distances from the observer to the source and the scatterer, respectively), $\theta_s\simeq \theta/x$ (for small angles) is the scattering angle, $f(x)$ is the dimensionless dust density distribution along
the LOS ($\int_0^1 f(x)\,dx =1$),
and $d\sigma_s(E,\theta_s)/d\Omega_s$ is the differential scattering cross section per one hydrogen atom, averaged over the dust grain distribution over sizes
and other grain properties (see, e.g., \citealt{1991ApJ...376..490M}). Here we assume that the source spectrum $F(E)$ does not vary: strong variability in the source can change the appearance of the halo in a complicated way, as time-lag depends upon radial distance from the source.

To understand the halo properties from simple analytical expressions, we will
use the Rayleigh-Gans (RG) approximation, in which the total scattering
cross section $\propto E^{-2}$; this approximation works better for higher
energies, $E \gtrsim 0.5$--2 keV, depending on the dust model.
For some dust models, the averaged differential cross section in the RG
 can be approximated as (Draine 2003)
\begin{equation}
\frac{d\sigma_s(E,\theta_s)}{d\Omega_s} \approx \frac{\sigma_s(E)}{\pi \theta_{s,50}^2}\,
\frac{1}{(1 + \theta_s^2/\theta_{s,50}^2)^2}\, ,
\end{equation}
where
\begin{equation}
\theta_{s,50} \approx \frac{\Theta}{E}\quad {\rm and} \quad
\sigma_s(E) \approx \frac{S}{E^2}\,10^{-22}\,{\rm cm}^2
\end{equation}
are the median scattering angle and the total cross section, respectively;
$E$ is the energy in keV.
The constant $\Theta$ in first eq.\ (A3) depends on the dust model; \citet{2003ApJ...598.1026D}
derived $\Theta = 360''$ from the dust model of \citet{2001ApJ...548..296W}, while \citet{2010A&A...520A..71B} found $\Theta = 7.4'$ for the model of \citet{1998ApJ...503..831S}.

It follows from the second eq. (A3) that the scattering optical depth is
\begin{equation}
\tau_{\rm scat}(E) \approx S N_{H,22} E^{-2}.
\end{equation}
The factor $S$ in the second eq.\ (A3) is a constant of the order of 1;
e.g., $S\approx 1.3$ from Figure 6 of \citet{2003ApJ...598.1026D}, while \citet{1995A&A...293..889P} found a mean value $S\approx 0.49$ for a number of halos observed with {\sl ROSAT} (but the scatter was very large), while \citet{1991ApJ...376..490M}
discuss models with $S= 0.903$, 1.09, and 0.47 (see their Table 1).
Costantini (2004; PhD thesis) estimated $\tau_{\rm sca}(1\, {\rm keV})$ for a number of halo sources observed with {\sl Chandra}; the values of $S$ derived from her results show a very strong scatter, $S$ from 0.018 to 2.26. The scatter itself may be natural, as the dust properties may be different for different sources.

It should be noted that the correlation of $\tau_{\rm sca}(1\,{\rm keV})$
with visual extinction $A_V$:
\begin{equation}
\tau_{\rm scat}(1\,\,{\rm keV}) =
(0.056 \pm 0.01) A_V
\end{equation}
\citep{1995A&A...293..889P} is better than that with $N_H$, but $A_V$ is rarely known for the objects
of interest.

Substituting (A2) and (A3) in (A1), we obtain the spectral intensity profile
\begin{equation}
I_{\rm halo}(\theta, E) = F(E) N_{H,22} \frac{S}{\pi \Theta^2} \int_0^1 \frac{f(x)}{x^2}\,
\left[1 + \left(\frac{\theta E}{x \Theta}\right)^2\right]^{-2} dx\,.
\end{equation}

For comparison with the point source + halo profile observed in the energy
 range
$E_1 < E < E_2$, the sum of the spectral intensities should be convolved with
the detector response, with allowance for the image spread caused by the
telescope and the detector. We have checked that the energy redistribution in the detector only slightly affects the broadband radial profile for a smooth incident spectrum. Therefore, assuming the observable halo size to be much larger than the PSF width, we obtain
\begin{equation}
I_{\rm obs}(\theta) = \int_{E_1}^{E_2} dE 
A_{\rm eff}(E)\, F(E)\,\left\{\psi(\theta,E) +
N_{H,22} \frac{S}{\pi \Theta^2}\int_0^1 dx \frac{f(x)}{x^2}\left[1 + \left(\frac{\theta E}{x\Theta}\right)^2\right]^{-2}\right\} ,
\end{equation}
where $A_{\rm eff}(E)$ is the detector's effective area, and $\psi(\theta,E)$ is the normalized PSF, which can be taken from a simulation (e.g., with MARX for {\sl Chandra} images). The first term in Equation (A7) corresponds to the point source, while the second term describes the halo.
 
The above equation can be integrated for a given set of halo parameters, and compared directly with the data. In particular, for the dust halo model shown in Figure 4 we picked 
 $\Theta=360\arcsec$, $S=1$, and the dust distribution  function
\begin{equation}
f(x)=
\begin{cases}
0 &\textrm{$x < x_1$,}\\
(x_2-x_1)^{-1} &\textrm{$x_1 \leq x \leq x_2$,}\\
0 &\textrm{$x > x_2$},
\end{cases}
\end{equation}
for $x_1=0.1$, $x_2=0.6$. According to (A6), this distribution corresponds to
the following profile of spectral intensity
\begin{equation}
I_{\rm halo}(\theta, E) = \frac{F(E) N_{H,22} S}{2\pi \theta^2 E^2 (y_2-y_1)}\left[\arctan\frac{y_2-y_1}{1+y_1y_2} - \frac{(y_2-y_1)(1-y_1y_2)}{(1+y_1^2)(1+y_2^2)}\right]\,,
\end{equation}
where $y_i=x_i\Theta/(\theta E)$. 
For small, intermediate, and large
$\theta$, Equation (A9) turns into
\begin{equation}
I_{\rm halo}(\theta, E) \simeq \frac{F(E) N_{H,22} S}{\pi\Theta^2}
\begin{cases}
(x_1x_2)^{-1} &   \theta\ll x_1\Theta/E, \\
(\pi/4x_2)(\Theta/\theta E) &  x_1\Theta/E \ll \theta \ll x_2\Theta/E, \\
(1/3)(\Theta/\theta E)^4 (x_1^2+x_1x_2+x_2^2) & \theta \gg x_2\Theta/E,
\end{cases}
\end{equation}
where the approximation for the intermediate $\theta$ implies
$x_2\gg x_1$.
It follows from Equation (A10) that, for the model (A8) (which may correspond 
to the case when the scattering occurs mainly in a Galactic arm),
the halo profile consists of three parts:
a flat top (whose size is
proportional to $x_1$), a slowly decreasing part ($\propto \theta^{-1}$), and
a steeply decreasing part ($\propto \theta^{-4}$), so that the characteristic size of a halo is about $x_2\Theta/E$.

\end{document}